\documentclass[a4paper,12pt]{article}

\usepackage{latexsym}

\newcommand{\nc}{\newcommand}    

\nc{\be}[1]{\begin{equation}\mbox{$\label{#1}$}}
\nc{\bea}[1]{\begin{eqnarray} \mbox{$\label{#1}$}}
\nc{\Section}[2]{\section{#2}\label{#1}}
\nc{\Bibitem}[1]{\bibitem{#1}}
\nc{\Label}[1]{\label{#1}}

\nc{\eea}{\end{eqnarray}}
\nc{\ee}{\end{equation}}

\nc{\bdm}{\begin{displaymath}}
\nc{\edm}{\end{displaymath}}
\nc{\dpsty}{\displaystyle}
\nc{\bc}{\begin{center}}
\nc{\ec}{\end{center}}
\nc{\ba}{\begin{array}}
\nc{\ea}{\end{array}}
\nc{\bab}{\begin{abstract}}
\nc{\eab}{\end{abstract}}
\nc{\btab}{\begin{tabular}}
\nc{\etab}{\end{tabular}}
\nc{\bit}{\begin{itemize}}
\nc{\eit}{\end{itemize}}
\nc{\ben}{\begin{enumerate}}
\nc{\een}{\end{enumerate}}
\nc{\bfig}{\begin{figure}}
\nc{\efig}{\end{figure}}

\nc{\arreq}{&\!=\!&}
\nc{\arrmi}{&\!-\!&}
\nc{\arrpl}{&\!+\!&}
\nc{\arrap}{&\!\!\!\approx\!\!\!&}
\nc{\non}{\nonumber}
\nc{\align}{\!\!\!\!\!\!\!\!&&}

\def\lsim{\; \raise0.3ex\hbox{$<$\kern-0.75em
      \raise-1.1ex\hbox{$\sim$}}\; }
\def\gsim{\; \raise0.3ex\hbox{$>$\kern-0.75em
      \raise-1.1ex\hbox{$\sim$}}\; }

\nc{\DOT}{\hspace{-0.08in}{\bf .}\hspace{0.1in}}
\nc{\Laada}{\hbox {$\sqcap$ \kern -1em $\sqcup$}}
\nc\loota{{\scriptstyle\sqcap\kern-0.55em\hbox{$\scriptstyle\sqcup$}}}
\nc\Loota{{\sqcap\kern-0.65em\hbox{$\sqcup$}}}
\nc\laada{\Loota}
\nc{\qed}{\hskip 3em \hbox{\BOX} \vskip 2ex}

\nc{\real}{{\rm I \! R}}
\nc{\Z}{{\sf Z \!\!\! Z}}
\nc{\complex}{{\rm C\!\!\! {\sf I}\,\,}}
\def\bigid{\leavevmode\hbox{\small1\kern-3.8pt\normalsize1}}
\def\id{\leavevmode\hbox{\small1\kern-3.3pt\normalsize1}}
\nc{\slask}{\!\!\!/}
\nc{\bis}{{\prime\prime}}
\nc{\pa}{\partial}
\nc{\na}{\nabla}
\nc{\ra}{\rangle}
\nc{\la}{\langle}
\nc{\goto}{\rightarrow}
\nc{\swap}{\leftrightarrow}

\nc{\EE}[1]{ \mbox{$\cdot10^{#1}$} }
\nc{\abs}[1]{\left|#1\right|}
\nc{\at}[2]{\left.#1\right|_{#2}}
\nc{\norm}[1]{\|#1\|}
\nc{\abscut}[2]{\Abs{#1}_{\scriptscriptstyle#2}}
\nc{\vek}[1]{{\rm\bf #1}}
\nc{\integral}[2]{\int\limits_{#1}^{#2}}
\nc{\inv}[1]{\frac{1}{#1}}
\nc{\dd}[2]{{{\partial #1}\over{\partial #2}}}
\nc{\ddd}[2]{{{{\partial}^2 #1}\over{\partial {#2}^2}}}
\nc{\dddd}[3]{{{{\partial}^2 #1}\over
    {\partial #2 \partial #3}}}
\nc{\dder}[2]{{{d #1}\over{d #2}}}
\nc{\ddder}[2]{{{d^2 #1}\over{d {#2}^2}}}
\nc{\dddder}[3]{{d^2 #1}\over
    {d #2 d #3}}
\nc{\dx}[1]{d\,^{#1}x}
\nc{\dy}[1]{d\,^{#1}y}
\nc{\dz}[1]{d\,^{#1}z}
\nc{\dl}[1]{\frac{d\,^{#1}l}{(2\pi)^{#1}}}
\nc{\dk}[1]{\frac{d\,^{#1}k}{(2\pi)^{#1}}}
\nc{\dq}[1]{\frac{d\,^{#1}q}{(2\pi)^{#1}}}

\nc{\bfT}{{\bf T }}

\def\TeV{{\rm\ TeV}}

\nc{\cA}{{\cal A}}
\nc{\cB}{{\cal B}}
\nc{\cD}{{\cal D}}
\nc{\cE}{{\cal E}}
\nc{\cG}{{\cal G}}
\nc{\cH}{{\cal H}}
\nc{\cL}{{\cal L}}
\nc{\cO}{{\cal O}}
\nc{\cT}{{\cal T}}
\nc{\cN}{{\cal N}}
\nc{\rvac}[1]{|{\cal O}#1\rangle}
\nc{\lvac}[1]{\langle{\cal O}#1|}
\nc{\rvacb}[1]{|{\cal O}_\beta #1\rangle}
\nc{\lvacb}[1]{\langle{\cal O}_\beta #1 |}
\nc{\bb}{\bar{\beta}}
\nc{\bt}{\tilde{\beta}}
\nc{\ctH}{\tilde{\cal H}}
\nc{\chH}{\hat{\cal H}}
\nc{\1}{\aa}
\nc{\2}{\"{a}}
\nc{\3}{\"{o}}
\nc{\4}{\AA}
\nc{\5}{\"{A}}
\nc{\6}{\"{O}}
\nc{\al}{\alpha}
\nc{\g}{\gamma}
\nc{\Del}{\Delta}
\nc{\e}{\textrm{e}}
\nc{\eps}{\epsilon}
\nc{\lam}{\lambda}
\nc{\Om}{\Omega}
\nc{\ve}{\varepsilon}
\nc{\mn}{{\mu\nu}}
\nc{\vp}{\varphi}

\nc{\advp}[3]{{\it  Adv.\ in\ Phys.\ }{{\bf #1} {(#2)} {#3}}}
\nc{\annp}[3]{{\it  Ann.\ Phys.\ (N.Y.)\ }{{\bf #1} {(#2)} {#3}}}
\nc{\apl}[3]{{\it  Appl. Phys. Lett. }{{\bf #1} {(#2)} {#3}}}
\nc{\apj}[3]{{\it  Ap.\ J.\ }{{\bf #1} {(#2)} {#3}}}
\nc{\apjl}[3]{{\it  Ap.\ J.\ Lett.\ }{{\bf #1} {(#2)} {#3}}}
\nc{\app}[3]{{\it Astropart.\ Phys.\ }{{\bf #1} {(#2)} {#3}}}
\nc{\cmp}[3]{{\it  Comm.\ Math.\ Phys.\ }{{ \bf #1} {(#2)} {#3}}}
\nc{\cqg}[3]{{\it  Class.\ Quant.\ Grav.\ }{{\bf #1} {(#2)} {#3}}}
\nc{\epl}[3]{{\it  Europhys.\ Lett.\ }{{\bf #1} {(#2)} {#3}}}
\nc{\ijmp}[3]{{\it Int.\ J.\ Mod.\ Phys.\ }{{\bf #1} {(#2)} {#3}}}
\nc{\ijtp}[3]{{\it Int.\ J.\ Theor.\ Phys.\ }{{\bf #1} {(#2)} {#3}}}
\nc{\jhep}[3]{{\it  JHEP\ }{{ \bf #1} {(#2)} {#3}}}
\nc{\jetpl}[3]{{\it  JETP Lett.\ }{{ \bf #1} {(#2)} {#3}}}

\nc{\jmp}[3]{{\it  J.\ Math.\ Phys.\ }{{ \bf #1} {(#2)} {#3}}}
\nc{\jpa}[3]{{\it  J.\ Phys.\ A\ }{{\bf #1} {(#2)} {#3}}}
\nc{\jpc}[3]{{\it  J.\ Phys.\ C\ }{{\bf #1} {(#2)} {#3}}}
\nc{\jap}[3]{{\it J.\ Appl.\ Phys.\ }{{\bf #1} {(#2)} {#3}}}
\nc{\jpsj}[3]{{\it J.\ Phys.\ Soc.\ Japan\ }{{\bf #1} {(#2)} {#3}}}
\nc{\lmp}[3]{{\it Lett.\ Math.\ Phys.\ }{{\bf #1} {(#2)} {#3}}}
\nc{\mpl}[3]{{\it  Mod.\ Phys.\ Lett.\ }{{\bf #1} {(#2)} {#3}}}
\nc{\ncim}[3]{{\it  Nuov.\ Cim.\ }{{\bf #1} {(#2)} {#3}}}
\nc{\np}[3]{{\it  Nucl.\ Phys.\ }{{\bf #1} {(#2)} {#3}}}
\nc{\pr}[3]{{\it Phys.\ Rev.\ }{{\bf #1} {(#2)} {#3}}}
\nc{\pra}[3]{{\it  Phys.\ Rev.\ A\ }{{\bf #1} {(#2)} {#3}}}
\nc{\prb}[3]{{\it  Phys.\ Rev.\ B\ }{{{\bf #1} {(#2)} {#3}}}}
\nc{\prc}[3]{{\it  Phys.\ Rev.\ C\ }{{\bf #1} {(#2)} {#3}}}
\nc{\prd}[3]{{\it  Phys.\ Rev.\ D\ }{{\bf #1} {(#2)} {#3}}}
\nc{\prl}[3]{{\it Phys\ Rev.\ Lett.\ }{{\bf #1} {(#2)} {#3}}}
\nc{\pl}[3]{{\it  Phys.\ Lett.\ }{{\bf #1} {(#2)} {#3}}}
\nc{\prep}[3]{{\it Phys\. Rep.\ }{{\bf #1} {(#2)} {#3}}}
\nc{\prsl}[3]{{\it Proc.\ R.\ Soc.\ London\ }{{\bf #1} {(#2)} {#3}}}
\nc{\ptp}[3]{{\it  Prog.\ Theor.\ Phys.\ }{{\bf #1} {(#2)} {#3}}}
\nc{\ptps}[3]{{\it  Prog\ Theor.\ Phys.\ suppl.\ }{{\bf #1} {(#2)} {#3}}}
\nc{\physa}[3]{{\it  Physica\ A\ }{{\bf #1} {(#2)} {#3}}}
\nc{\physb}[3]{{\it  Physica\ B\ }{{\bf #1} {(#2)} {#3}}}
\nc{\phys}[3]{{\it Physica\ }{{\bf #1} {(#2)} {#3}}}
\nc{\rmp}[3]{{\it  Rev.\ Mod.\ Phys.\ }{{\bf #1} {(#2)} {#3}}}
\nc{\rpp}[3]{{\it Rep.\ Prog.\ Phys.\ }{{\bf #1} {(#2)} {#3}}}
\nc{\sjnp}[3]{{\it Sov.\ J.\ Nucl.\ Phys.\ }{{\bf #1} {(#2)} {#3}}}
\nc{\spjetp}[3]{{\it Sov.\ Phys.\ JETP\ }{{\bf #1} {(#2)} {#3}}}
\nc{\yf}[3]{{\it Yad.\ Fiz.\ }{{\bf #1} {(#2)} {#3}}}
\nc{\zetp}[3]{{\it Zh.\ Eksp.\ Teor.\ Fiz.\  }{{\bf #1}  {(#2)} {#3}}}
\nc{\zp}[3]{{\it Z.\ Phys.\ }{{\bf #1} {(#2)} {#3}}}
\nc{\ibid}[3]{{\sl ibid.\ }{{\bf #1} {#2} {#3}}}

\nc{\rf}[1]{(\ref{#1})}
\nc{\nn}{\nonumber \\*}
\nc{\bfB}{\bf{B}}
\nc{\bfv}{\bf{v}}
\nc{\bfx}{\bf{x}}
\nc{\bfy}{\bf{y}}
\nc{\vx}{\vec{x}}
\nc{\vy}{\vec{y}}
\nc{\oB}{\overline{B}}
\nc{\oI}{\overline{I}}
\nc{\oR}{\overline{R}}
\nc{\rar}{\rightarrow}
\nc{\ti}{\times}
\nc{\slsh}{\hskip-5pt/}
\nc{\sm}{Standard~Model~}
\nc{\MP}{M_{\rm Pl}}
\nc{\tp}{t_{\rm Pl}}

\nc{\pmin}{p_{\rm min}}
\nc{\pmax}{p_{\rm max}}
\nc{\fo}{f_0}
\nc{\foi}{f_{0,i}\,}
\nc{\fop}{f_0^P}
\nc{\fou}{f_0^U}

\nc{\eff}{{\rm eff}}
\nc{\MT}{M_{\rm T}}
\nc{\ML}{M_{\rm L}}
\nc{\kk}{\vek{k}}
\nc{\pp}{{\rm p}}
\nc{\pt}{\partial_t}
\nc{\w}{\omega}
\nc{\uhat}{\hat{U}_\w}

\nc{\etal}{\mbox{\it et al.}}
\nc{\ie}{{\it i.e. }}
\nc{\eg}{{\it e.g. }}
\nc{\trh}{T_{\rm RH}}
\nc{\ad}{{a'\over a}}
\nc{\bd}{{b'\over b}}
\nc{\Rd}{{R'\over R}}
\nc{\hyp}{\,\; F_{1{\hskip -16pt}2}{\hskip 11pt}}
\nc{\half}{{1\over 2}}

\begin{document}
{\title{\vskip-2truecm{\hfill {{\small UB-ECM-PF 03/01\\
   }}\vskip 1truecm}
{\LARGE Cosmological solutions of braneworlds with warped and
  compact dimensions}}
\vspace{-.2cm}
{\author{
{\sc Tuomas Multam\"aki$^{1}$}
\\
{\sl\small 
Departament Estructura i Constituents de la Materia,}\\
{\sl\small Universitat de Barcelona, Diagonal 647, 08028 Barcelona, Spain}\\
{\sl and}\\
{\sc Iiro Vilja$^{2}$ }
\\
{\sl\small Department of Physics, University of Turku, FIN-20014 Turku, Finland}\\
}
\date{ }
\maketitle
\begin{abstract}
\noindent
We study cosmological aspects of braneworld models with a warped
dimension and an arbitrary number of compact dimensions. With a
stabilized radion, a number of different cosmological bulk
solutions are found in a general case. Both one and
two brane models are considered. The Friedmann equation is
calculated in each case. Particular attention is paid to six dimensional models
where we find that the usual Friedmann equation can typically be
recovered without fine-tuning.

\end{abstract}
\vfill
\footnoterule
{\small $^1$tuomas@ecm.ub.es\\
{\small $^2$vilja@utu.fi}

\thispagestyle{empty}
\newpage
\setcounter{page}{1}

\section{Introduction}
The study of extra dimensions has \cite{rubakov}, in addition to the novel features
of particle phenomenology, inspired much research on the cosmological
aspects of extra dimensional models. Especially the study of
brane cosmology has recently been an active field \cite{langlois}. 
Already very soon after the pioneering papers on brane world models
\cite{rs}, the peculiar cosmological aspect of these constructions
were realized \cite{bdl}. Later on, a number of papers have studied
the cosmological bulk solutions of RS-type models,  \cite{bulksols}-\cite{kantiprd62}.

An important aspect of any cosmological brane world model is the
question of radion stabilization, which in the RS-model corresponds
to the stabilization of the distance between the two three branes.
At the same time one wishes to solve the hierarchy problem introducing
a fundamental, D -dimensional  gravity scale $M_*$ which is related
suitably to the Planck mass $M_{Pl}$, at least at the present time.
Most interesting would be if the fundamental scale $M_*$ is near
1 TeV and hence in reach of the future colliders.

In addition to the many studies on ADD \cite{add} and RS \cite{rs} -models, a number of
papers have combined the two so that in addition to the warped
dimension, a compact extra dimension is also present, see \eg 
\cite{misha}-\cite{me}. The stabilization of six
dimensional models have been discussed in \cite{burgess, kanti}.
Combining the two models can be seen as a quite natural scenario since
in any case we expect to have a number compact dimensions coming from
string theory. Obviously these compact dimensions do not have to
be 'large' but such a possibility is worth studying.
Thus of the D-1 spatial dimensions three are the usual observed dimensions,
one is a Randall-Sundrum like dimension and D-5 are compact, but unconstrained
dimensions. Hence the dimension of the brane is  D-2. Because the
Standard Model particles are free to propagate in the extra compact 
dimensions, their size are constrained by experiments.
In particular in our models, which have the extra dimension topology
of $I\times (S^1)^{(D-5)}$, where $I\subseteq \real$,
the radii of $S^1$ factors are constrained to be smaller than 
about 1 TeV$^{-1}$.

In this paper we are interested in the cosmological
properties of brane world models, where the number of the extra  dimensions
is two or more. It is organized as follows: First we study the
D-dimensional ($D\geq 5$) 
Einstein's equations in order to find some cosmological bulk solutions. Writing
the D-dimensional jump conditions we solve the Friedmann
equation in each case. Then we concentrate on the 6-dimensional models and 
their properties and finally present conclusions.

\section{Einstein's equations in D-dimensions}
In order to write the Einstein's equations in D-dimension, we first 
define the coordinates\footnote{
In this paper we follow a convention where capital roman letters
denote the full D-dimensional space-time, small roman letters 
the three dimensional
space and small greek letters the four dimensional space-time, \eg 
$A=0,...,D-1,\ a=1,2,3,\ \alpha=0,...,3$.} as:
\be{coords}
x_{\mu}=(t,x_1,x_2,x_3,z,\theta_1,...,\theta_{D-5}).
\ee
Due to the symmetries, the metric tensor can always be chosen to have  diagonal
form with $b_i=b_i(t,z)$.
Hence the line element can be written as
\be{genline}
ds^2=\eta(t,z)^2dt^2-R(t,z)^2dx_idx^i-a(t,z)^2dz^2-\sum_{i=1}^{D-5}b_i(t,z)^2d\theta_i^2.
\ee
With this form of the metric tensor, the Einstein's equations can be written compactly as
\be{eesequs}
G_{AB}=-\kappa^2 T_{AB},
\ee
where $\kappa$ is the $D$-dimensional gravitational constant,
$G_{AB}$ is the Einstein's tensor in $D$-dimensions and 
$T_{AB}$ is the $D$-dimensional energy-momentum tensor. It includes
both the bulk and the brane energy-momentum tensors. In the bulk the
energy-momentum tensor is assumed to have the form 
\be{bulktmunu}
T_A^B={\textrm {diag}}(-\Lambda_B,-\Lambda_B,-\Lambda_B,-\Lambda_B,
T^4_4(t,z),...,T^{D-1}_{D-1}(t,z)).
\ee 
In other words, we have a normal
cosmological constant in the bulk but the cosmological constants in
the compact dimensions are allowed to be functions of time and the
$z$-coordinate. As it was already seen in \cite{kanti1}, this is necessary in
order to satisfy the continuity (and Einstein's) equations in the bulk.

We now look for solutions with a stable radion field, {\it i.e.}
$\dot a = 0$. We are free to choose it to have a constant value in
$z$-direction by redefinition of $z$-coordinate, thus
simply $a(t,z)=1$.
We assume further that the compact dimensions are also stabilized, 
$b_i(t,z)=b_i(z)$.
Non-zero components of Einstein's tensor are now:
\bea{ig00}
G_0^0 & = & 
\sum_{i\neq j\geq 1}^{D-5}{b_i'b_j'\over b_i b_j}+
\sum_{i=1}^{D-5}{b_i''\over b_i}+
3\sum_{i=1}^{D-5}{b_i'\over b_i}{R'\over R}+
 3\frac{{R'}^2}{\,{R}^2} + 
  3\frac{R''}{\,R} - 
  3\frac{{\dot{R}}^2}{{\eta}^2\,{R}^2}\nonumber\\
G_l^l & = &   
\sum_{i\neq j\geq 1}^{D-5}{b_i'b_j'\over b_i b_j}+
\sum_{i=1}^{D-5}{b_i''\over b_i}+
\sum_{i=1}^{D-5}{b_i'\over b_i}{\eta'\over\eta}+
2\sum_{i=1}^{D-5}{b_i'\over b_i}{R'\over R}+
2 \frac{\eta'\,R'}{\,\eta\,R}+ 
  \frac{{R'}^2}{\,{R}^2} +\nonumber\\
& &   \frac{\eta''}{\,\eta} +
  2\frac{R''}{\,R} +
  2\frac{\dot{\eta}\,\dot{R}}{{\eta}^3\,R} - 
  \frac{{\dot{R}}^2}{{\eta}^2\,{R}^2} -
  2\frac{\ddot{R}}{{\eta}^2\,R}\ (l=1,2,3)\nonumber\\
G_4^4 & = & 
\sum_{i\neq j\geq 1}^{D-5}{b_i'b_j'\over b_i b_j}+
{\eta'\over\eta}\sum_{i=1}^{D-5}{b_i'\over b_i}+
 3{R'\over R}\sum_{i=1}^{D-5}{b_i'\over b_i}+
 3\frac{\,\eta'\,R'}{\,\eta\,R}+ 
 3 \frac{\,{R'}^2}{\,{R}^2}+
 3 \frac{\,\dot{\eta}\,\dot{R}}{{\eta}^3\,R}- 
 3 \frac{\,{\dot{R}}^2}{{\eta}^2\,{R}^2}-
 3 \frac{\,\ddot{R}}{{\eta}^2\,R}\\
G_n^n & = & 
\sum_{i\neq j\neq n}^{D-5}{b_i'b_j'\over b_i b_j}+
\sum_{n\neq i}^{D-5}{b_i''\over b_i}+
({\eta'\over\eta}+3{R'\over R})\sum_{n\neq i}^{D-5}{b_i'\over b_i}+
3\frac{\eta'\,R'}{\,\eta\,R}+\frac{3\,{R'}^2}{\,{R}^2}+
  \frac{\eta''}{\,\eta}+\nonumber\\
& &   3\frac{\,R''}{\,R}+
  3\frac{\,\dot{\eta}\,\dot{R}}{{\eta}^3\,R}-
  3\frac{\,{\dot{R}}^2}{{\eta}^2\,{R}^2}- 
  3\frac{\,\ddot{R}}{{\eta}^2\,R}\ (n=5,...,D-1)\nonumber\\
G_{04} & = & -3\frac{\eta'\,\dot{R}}{\eta\,R}+3\frac{\dot{R}'}{R}\nonumber
\eea
As we see, the Einstein's tensor does have a non-diagonal component $G_{04}$
whereas the  energy-momentum tensor is diagonal. Thus as the first step we solve
The equation $G_{04}=0$ by writing  $\eta(t,z)$ in the form
\be{etasol}
\eta(t,z)=\lambda(t) \dot{R}(t,z),
\ee
where $\lambda$ is a arbitrary positive time dependent function. 
This is naturally the same solution that was found in the 
5-dimensional case \cite{kanti1,kantiprd62}.

Using the solution (\ref{etasol}) for $\eta(t,z)$, the non-zero components of  
$G_A^B$ become rewritten as
\bea{g00}
G^0_0 & = & -{3\over \lambda^2 R^2}+
\sum_{i\neq j\geq 1}^{D-5}{b_i'b_j'\over b_i b_j}+
\sum_{i=1}^{D-5}{b_i''\over b_i}+
3\sum_{i=1}^{D-5}{b_i'\over b_i}{R'\over R}+
3 ({R'\over R})^2+3{R''\over R}\nonumber\\
G^l_l & = & -\frac{1}{{{\lambda}}^2\,{R}^2}+ 
\sum_{i\neq j\geq 1}^{D-5}{b_i'b_j'\over b_i b_j}+
\sum_{i=1}^{D-5}{b_i''\over b_i}+
(2{R'\over R}+{\dot{R'}\over\dot{R}})\sum_{i=1}^{D-5}{b_i'\over
b_i}+\frac{R'^2}{R^2}+\nonumber\\
& & 
2\frac{\,R''}{R}+2\frac{\,\dot{\lambda}}{\lambda^3\,R\,\dot{R}}+
2\frac{\,R'\,\dot{R}'}{R\,\dot{R}}+\frac{\dot{R}''}{\dot{R}}\ (l=1,2,3)\\
G^4_4 & = & 
-\frac{3}{{{\lambda}}^2\,{R}^2} + 
\sum_{i\neq j\geq 1}^{D-5}{b_i'b_j'\over b_i b_j}+
(3{R'\over R}+{\dot{R'}\over\dot{R}})\sum_{i=1}^{D-5}{b_i'\over b_i}+
3\frac{\,{R'}^2}{{R}^2} + 
3\frac{\,\dot{\lambda}}{{{\lambda}}^3\,R\,\dot{R}} + 
3\frac{\,R'\,\dot{R}'}{R\,\dot{R}}\nonumber\\
G^n_n & = & -\frac{3}{{{\lambda}}^2\,{R}^2} + 
\sum_{i\neq j\neq n}^{D-5}{b_i'b_j'\over b_i b_j}+
\sum_{n\neq i}^{D-5}{b_i''\over b_i}+
(3{R'\over R}+{\dot{R'}\over\dot{R}})\sum_{n\neq i}^{D-5}{b_i'\over b_i}+
3\frac{\,{R'}^2}{{R}^2} + 
3\frac{\,R''}{R}+\nonumber\\
& & 3\frac{\,\dot{\lambda}}{{{\lambda}}^3\,R\,\dot{R}} + 
3\frac{\,R'\,\dot{R}'}{R\,\dot{R}} + 
\frac{\dot{R}''}{\dot{R}}\ (n=5,...,D-1).\nonumber
\eea
The Einstein's equations which are directly related to the
4-dimensional space-time properties via couplings to 4-dimensional
matter are the equations for $G^0_0$ and $G^l_l$.
In contrast the other components of the Einstein's tensor are related to the unknown
components $T_4^4,\ ...,T_{D-1}^{D-1}$ of  the  energy-momentum tensor and remain,
at this stage, less constrained.

Besides the Einstein's equations,  the equations to be satisfied include also
the continuity equations in D-dimensional space-time, $T^{AB}_{\ \ \ ;B}=0$. However, in the
bulk only one of these continuity equations is non-trivial giving
\be{bulkcont}
T^{4 B}_{\ \ \ ;B}=T^{4'}_4+(T_4^4-\Lambda_B)
(3{R'\over R}+{\dot{R'}\over\dot{R}})+
\sum^{D-1}_{i=4}(T_4^4-T^i_i){b_i'\over b_i}=0.
\ee
We remind the reader, that the continuity equations are automatically 
satisfied due to  Bianchi identities whenever the Einstein's equations
have been solved. 

\section{Cosmological bulk solutions}\label{bulks}
To reach our goal to study the cosmological properties of the brane models
we have to solve first the Einstein equations (\ref{eesequs}) in the bulk. The whole
set of equations nor even the first four are hardly solvable in
general. However, our special choices allows us to relate these to each other.

Thus in order to solve the Einstein equations we simplify the
set of equations (\ref{g00}) further.
We first note that since $\eta(t,z)$ is given by (\ref{etasol}),
we can write $G_i^i$ in terms of $G_0^0$:
\be{genegii}
G_i^i=G_0^0+{1\over 3}{R\over\dot{R}}\partial_t G_0^0\ \ {\rm (no\ sum)},
\ee
hence in the bulk, as long as $\Lambda_B$ is constant, we only need to look for solutions of the first 
Einstein's equation $G_0^0=\kappa^2\Lambda_B$ in order to solve the
first four Einstein's equations.

Even now,
the general solution of the equation $G_0^0=\kappa^2\Lambda_B$  is not known. We can,
however, try to solve it in  two simple special cases: We can use either 
exponential {\it Ans\"atze} where it is assumed that $b_i(z)=\exp(k_i z)$ or 
a $\cosh $ -{\it Ansatz}, which we call the bowl-model, where 
$b_i(z)=b_i(0)\cosh^{2\over d+1}(k z)$ and $d=D-5$ is the number of 
compact extra dimensions.
These two choices carry some essential and interesting properties as will be seen. 
We first study the exponential model and then the bowl-model.

\subsection{Exponential models}
We first assume the exponential {\it Ans\"atze} for
the scale factors of the compact dimensions, $b_i(z)=b_i(0)\exp(k_i z)$. 
The $(00)$-component of the Einstein's equations in the bulk is then
\be{geneg00}
-{3\over\lambda^2 R^2}+\sum_{i\geq j=1}^{D-5}k_ik_j+3{R'\over R}\sum_{i=1}^{D-5}
k_i+3({R'\over R})^2+3{R''\over R}=\kappa^2\Lambda_B,
\ee
which is easily solvable. The nature of the solution depends, however, on the relations
between the parameters of the equation.
Indeed, there are four distinct classes of solutions, in addition to the trivial case
$k_i=0\ \forall i$, to equation Eq. (\ref{geneg00}),
depending on the values of the constants $k_i$. 

By defining the
combinations of the constants $k_i$ as $A=\sum_{i\geq j=1}^{D-5}k_ik_j$ and
$B=\sum_i^{D-5}k_i$ the solutions to 
Eq. (\ref{geneg00}) are given by
\bea{genesols}
{\bf A)}\ B^2\ne {8\over 3}(A-\kappa^2\Lambda_B)\ne 0:\ R^2(t,z) & = &
3/(\lambda^2(t)(A-\kappa^2\Lambda_B))+\nonumber\\
& & d_1(t)\exp\Big({z\over 6}(-3B+\sqrt{9B^2-24(A-\kappa^2\Lambda_B)})\Big)+\nonumber\\
& & d_2(t)\exp\Big({z\over
6}(-3B-\sqrt{9B^2-24(A-\kappa^2\Lambda_B)})\Big)\nonumber\\
{\bf B)}\ A=\kappa^2\Lambda_B,\ B\ne 0: \ \ \ \ \ \ \ \ R^2(t,z)&= & 2z/(\lambda^2(t)B)+
d_1(t)\exp(-B z)+d_2(t)\\
{\bf C)}\ A=\kappa^2\Lambda_B,\ B=0: \ \ \ \ \ \ \ \ R^2(t,z)&= & d_1(t)+d_2(t)z+z^2/\lambda^2(t)\nonumber\\
{\bf D)}\  B^2={8\over 3}(A-\kappa^2\Lambda_B)\neq 0: \ R^2(t,z)&= & 3/(\lambda^2(t)
(A-\kappa^2\Lambda_B))+\exp(-{B\over 2}z)\Big(d_1(t)+d_2(t)z\Big).\nonumber
\eea
Any of these solutions will satisfy the first four Einstein's
equations.
In order that the rest of the Einstein's equations are also valid in
the bulk, 
we must generally have non-trivial components of the energy momentum
tensor \ie $T_i^i=T_i^i(t,z),\ i=4,...,D-1$.

\subsection{Bowl-models}
Of particular interest are models with only a single brane
with matter, hence one does not have to be concerned with the
properties and implications of matter on a hidden brane.
This means that either we have no second brane at all or it only has
tension and no matter. Clearly, if we wish to have only a single
brane, we must have more than one warped extra dimension so that
we can wrap the brane along a compact extra dimension. The 
coordinate point that we have to be concerned with in such a
construction is the origin, $z=0$. Having a vanishing stress tensor at
$z=0$ means that the jumps of $R$ and $b_i$ must vanish
at the origin \ie the first derivative of the functions must be
continuous.
Clearly, the exponential {\it Ans\"atze} for the $b_i$:s do not
satisfy these conditions and 
in general it is not easy to see whether solutions where 
$R'(t,0)=b_i'(t,0)=0$, exist or not. 

As a special case we can construct a solution by 
choosing an essentially common {\it Ansatz} for the compact
extra dimensions:
\be{genbowlans}
b_i(z)=b_i(0)\cosh^{2\over d+1}(k z).
\ee
With this {\it Ansatz}, we can solve for $R(z,t)$:
\bea{genbowlsol}
R^2(t,z) & = & {2\over\beta\lambda^2}\nonumber\\
& & + d_1(t) e^{(z/2(\alpha-\sqrt{\alpha^2-4\beta}))}\hyp
({\alpha-\sqrt{\alpha^2-4\beta}\over 2k},{\alpha\over
2k},1-{\sqrt{\alpha^2-4\beta}\over 2k},e^{2 k z})\\
& & + d_2(t)  e^{(z/2(\alpha+\sqrt{\alpha^2-4\beta}))}\hyp
({\alpha+\sqrt{\alpha^2-4\beta}\over 2k},{\alpha\over
2k},1+{\sqrt{\alpha^2-4\beta}\over 2k},e^{2 k z}),\nonumber
\eea
where $\alpha={2dk\over d+1}$,
$\beta={2\over 3}(\alpha k-\kappa^2\Lambda_B)$ and 
$\hyp(a,b;c;z)$ is the hypergeometric function. 
One can now choose one of the $d_i(t)$'s in such a way that $R'(0)$
vanishes. 

Unfortunately, this solution is not very intuitive, and so, in
order to study the properties of the cosmological bowl solutions,
we restrict ourselves to the case $k^2=2 \frac {d+1}{d+3} \kappa^2\Lambda_B$.
In six dimensions ($d=1$)  this implies 
\be{sixdbowl2}
R^2(t,z)={2\over\lambda^2(t)\kappa^2\Lambda_B}\ln(\cosh(k z))+d(t),
\ee
whereas for $d>1$
we write
\be{morethansixdbowl2}
R^2(t, z)= d_1(t)\cosh^\nu (kz) - {2\over \lambda^2(t)k^2\nu},
\ee
where the exponent is given by $\nu = 1-{2d\over d+1}$. Note, that
the condition $R'(0)=0$ has already been exploited
so that there is only one brane in the scenario which, technically, removes
one of arbitrary functions $d_1, d_2$ of the Eq. (\ref{genbowlsol}). Note also, that
these solutions correspond necessarily to a positive bulk cosmological constant.
This model will be studied in more detail in Section \ref{6dcase}.

\subsection{Jump equations}
Next we move forward to study the properties of the models on the brane(s).
We relate the energy momentum tensor of the D-1 -dimensional brane to the 
solutions of D-dimensional Einstein equations. We idealize the brane to be a
zero width object in $z$-direction regardless how the brane has been formed.
Thus, any brane locates at an incontinuity of the $z$-derivative of the metric tensor, 
in particular at the edge of the space.

So, the energy content of the brane is related to the global solutions
through the jump equations \cite{bdl}. The jump equations in the D-dimensional 
case can be read from the
Einstein's equations as usual. The brane contribution to the energy-momentum
tensor is supposed to be confined to some $z=z'$, {\it i.e.} it is assumed
to have the form 
\be{genebrane}
(T_A^B)_{Br} = \delta(z-z')\,
\tilde T_A^B=\delta(z-z')\,\textrm{diag}(\Lambda+\rho,\Lambda-p,\Lambda-p,\Lambda-p,0,
\Lambda_1-P_1,...,\Lambda_{D-5}-P_{D-5}).
\ee
Thus there is a uniform matter distribution also in the directions of
the compact dimensions.
Solving for the different components we get (for general $b_i=b_i(t,z))$:
\bea{genejumps}
{[R']\over R}\Bigg|_{z'} & = & {\kappa^2\over D-2}\Big((D-5)(\rho+p)-(D-6)(\Lambda+\rho)+
\sum_{i=1}^{D-5}(\Lambda_i-P_i)\Big)\nonumber\\
{[\eta']\over\eta}\Bigg|_{z'} & = & {[R']\over R}-\kappa^2(\rho+p)\\
& = & {\kappa^2\over D-2}\Big(-3(\rho+p)-(D-6)(\Lambda+\rho)+
\sum_{i=1}^{D-5}(\Lambda_i-P_i)\Big)\nonumber\\
{[b_j']\over b_j}\Bigg|_{z'} & = & {\kappa^2\over 2(D-2)}\Big(4\Lambda+\rho-3p+
\sum_{i=1,\ i\ne j}^{D-5}(\Lambda_i-P_i)+(3-D)(\Lambda_j-P_j)\Big)\nonumber
\eea
From these, we see that there are two special cases, $D=5$ where
there is no pressure ($p$) (nor compact dimensions)
dependence on the jump of $R$, and $D=6$, where
the brane tension $\Lambda$ does not contribute either to the jump
of $R$ or $\eta$.

One of  the continuity equations  ($\tilde T^{0A}_{\ \ \ ;A}=0$),  
takes on the brane the usual form
\be{contbrane}
\dot{\rho}+3H(\rho+p)=0,
\ee
where $H=\dot{R}/R|_{z'}$. Because we have the freedom to choose $\lambda$,
we may use the standard time at a brane by  taking $\lambda (t) = 1/\dot R(t, z')$.
Note, however, that if there is more than one brane, standard time can be chosen
only at one of these.
The other constraints can be obtained by noting that from the solution of $\eta$, it follows that
$\eta'/\eta=R'/R+R/\dot{R}\partial_t(R'/R)$. On the brane, this relates the jumps by
\be{genejumpcond}
{[\eta']\over\eta}\Bigg|_{z_0}={[R']\over R}\Bigg|_{z_0}+
{1\over H}\partial_t\Big({[R']\over R}\Bigg|_{z_0}\Big).
\ee
Using the jump conditions and the continuity equation
(\ref{contbrane}), we get a relation that binds the brane energy density, $\rho$, and
brane pressures, $p,\ P_i$ by
\be{geneenergy}
(D-5)(\dot{p}-{1\over 3}\dot{\rho})-\sum_{i=1}^{D-5}\dot{P_i}=0.
\ee
Hence, if the jump condition for $R'$ is satisfied 
and Eq. (\ref{geneenergy}) holds, the jump condition for
$\eta'$ is automatically satisfied.
Note that in cosmological 5D-models \cite{bulksols,kanti1},
relation (\ref{genejumpcond}) reduces to the usual continuity equation
and hence a solution which satisfies the jump condition for $R'$
will also satisfy the condition for $\eta'$, which was noted
in \cite{kantiprd62}.
It is clear that in higher dimensional models, some energy can be 
flowing in/out of the compact dimensions when the universe is
not radiation dominated ($p\neq\rho/3$). 

By assuming that the compact dimensions are static, $\dot b_j=0$,
we get by differentiating the jump equation for $b'_j/b_j$ with 
respect to time, an equation
for each $j=1,...,\ d$ that relates the energy and pressures in different 
dimensions on the brane,
\be{bcont}
\dot{\rho}-3\dot{p}-\sum_{i=1,\ i\ne j}^{D-5}\dot{P}_i-(3-D)\dot{P}_j=0.
\ee
Note that these equations and Eq. (\ref{geneenergy})
are not linearly independent, so that the system
is not overconstrained. By summing the equations for each $j$, we 
get again Eq. (\ref{geneenergy}) and hence we only have $d$ equations
for $d$ unknowns.

\subsection{Friedmann equations}
With the global bulk solutions and the jump conditions, we can now
calculate the Friedmann equation on the brane in each case. 
At first we shall be concerned with models with two branes 
separated by distance $L$, which may be negative.
The branes are located at $z=z_0$ and $z=z_1=z_0+L$.
We normalize the scale factor $R(t,z)$ so that $R(t,z_0)=a_0(t)$ \ie
we live on the brane at $z_0$.
The Friedmann equation in each case can be calculated by first
calculating the jump of the first derivative at the two branes,
which allows us to express $\lambda(t)$ in 
terms of $\sigma_i\equiv -[R']/R|_{z_i}$:s. 
Since the Hubble constant on our brane is given by
$H^2=(\dot{a_0}/a_0)^2={1/(a_0^2\lambda^2)}$,
we can then express $H$ as a function of $\sigma_i$:s.

The Friedmann equations for the four types of the exponential models
(\ref{genesols}) are:
\bea{genefrieds}
{\bf A)}\ H^2 & = & {A+\kappa^2\Lambda_B\over 3}
\frac{B^2+
    4\,B\,
     \left( {\sigma_0} + {\sigma_1} \right)  - 
    4\,\left(\gamma^2 + 
       2\, {(e^{2\,\gamma L}+1)\over(e^{2\gamma L}-1)} \,\gamma
        \left( {\sigma_0} - {\sigma_1} \right)  - 
       4
        {\sigma_0}\,{\sigma_1} \right) }{
    B^2+ 
      4\,B\,
       {\sigma_1} - 
      4\,\gamma \left( \gamma - 
         2\,\left( (e^{2\,\gamma L}+1)- 
            2\,e^{\frac{\left( B + 2\gamma \right) \,L}{2}}
            \right) \,{\sigma_1}(e^{2\gamma L}-1)^{-1} \right)}\nonumber\\
{\bf B)}\ H^2 & = & 
B {2(e^{B L}-1)\sigma_0\sigma_1-B(\sigma_0-\sigma_1 e^{BL})\over
2(e^{B L}-1)\sigma_1+B(e^{B L}-1-2\sigma_1 L e^{B L})}\\
{\bf C)}\ H^2 & = & {2\sigma_0\sigma_1L-\sigma_0+\sigma_1\over L(1-\sigma_1L)}\nonumber\\
{\bf  D)}\  H^2 & = & {B^2\over 8}
{B^2L+4\sigma_1(2+BL)+4\sigma_0(BL-2+4\sigma_1L)\over B^2L+
4\sigma_1(2+BL-2 e^{BL/2})},\nonumber
\eea
where $\gamma=\sqrt{9B^2-24\,(A-\kappa^2\Lambda_B)}$.
By using the jump equations, one can write these in terms
of the quantities defined on the branes. 

For the bowl-model the Friedmann equation has a remarkably simple form. We obtain
\be{bowlfried}
H^2= \frac k2 [2\sigma_0\coth (kz_0) -k\nu],
\ee
which is valid for all $d\geq 1$.

We study these equations in more detail in the next Section,
where the six-dimensional case is considered. Generalization to
higher dimensions can then be done straightforwardly.


\section{Six-dimensional case}\label{6dcase}
As we have seen from the general expressions, in six dimensions
there is no contribution to the jumps of $R$ or $\eta$ from the
brane tension. Also, with only one compact dimension, the extra dimension can be identified
with $\real^2$ and we can
write the cosmological solution for a model with only a single
brane wrapped around the origin with no additional matter or
tension carrying brane. In this Section we study these properties 
in more detail.

The $(00)$-component of the Einstein's equations in six dimension,
after substituting $\eta(t,z)=\lambda(t)\dot{R(t,z)}$, is
\be{6d00eq}
-{3\over \lambda^2 R^2}+{b''\over b}+
3{b'\over b}{R'\over R}+3 ({R'\over R})^2+3{R''\over R}=
\kappa^2\Lambda_B.
\ee

Let us first look for solutions with the exponential {\it Ansatz},
$b(t,z)=b_0\exp(k z)$. Again, depending on the value of the parameters
we have different types of non-trivial solutions, which can be read out from
Eqs. (\ref{genesols}). However, the case {\bf C} is possible only if $k=\Lambda_B= 0$,
corresponding constant metric component $b$ and the solution given in Eq. (\ref{genesols}).
Thus, because  $A=B^2=k^2$,  the cases that remain are:

\vspace*{0.2cm}
\noindent {\bf A)} $k^2\neq \kappa^2\Lambda_B$, $k^2\neq 8\kappa^2\Lambda_B/5$:
\bea{e0s1}
R^2(t,z) & = & d_1(t)\exp(({z\over 6}(-3k+\sqrt{24\kappa^2\Lambda_B-15 k^2}))+\nonumber\\
& & d_2(t)\exp({z\over 6}(-3k-\sqrt{24\kappa^2\Lambda_B-15k^2}))+
{3\over (k^2-\kappa^2\Lambda_B)\lambda^2}\nonumber
\eea
\noindent {\bf B)} $k^2=\kappa^2\Lambda_B\ne 0$:
\be{e0s2}
R^2(t,z)=d_1(t)e^{-k z}+d_2(t)+{2z\over k\lambda^2}
\ee
\noindent {\bf D)} $k^2=-8\kappa^2\Lambda_B/5\ne 0$:
$$
R^2(t,z)=-{5\over \lambda^2(t)\kappa^2\Lambda_B}+\exp(-{k\over 2}z)\Big(d_1(t)+d_2(t)z\Big)
$$

In addition to these, we also have the bowl-solutions for which $b'(0)$
vanishes. In six dimensions the {\it Ansatz} for $b$ is $b(z)=\cosh (kz)$ and generally we write
\bea{6bowl1}
R^2(t,z) & = & d_1(t)\exp({kz\over 2}(1-\alpha))
\hyp (\half,\half(1-\alpha),1-\half\alpha,-e^{2kz})+\nonumber\\
 & & d_2(t)\exp({kz\over 2}(1+\alpha))\hyp (\half,\half(1+\alpha),1+\half\alpha,-e^{2kz})+
{3 \over \lambda^2(k^2-\kappa^2\Lambda_B)},
\eea
where $\alpha\equiv\sqrt{1-{8\over 3}(1-{\kappa^2\Lambda_B\over k^2})}$ and 
$\hyp(a,b;c;z)$ is the hypergeometric function. In particular, if  $k^2=\kappa^2\Lambda_B$
we obtain the special solution given by Eq. (\ref{sixdbowl2}).

\subsection{Jumps}
The cosmological solutions obtained by placing branes at $z=z_0$ and
$z=z_1$ can now be studied as previously. From the jump equations 
(\ref{genejumps}) we get:
\bea{6dumps}
{[b']\over b}\Bigg|_{z_i} & = & {\kappa^2\over 4}(4\Lambda_i+\rho_i-3p_i-
3\Lambda_i^\theta+3P_i)\nonumber\\
{[R']\over R}\Bigg|_{z_i} & = & {\kappa^2\over 4}(\rho_i+p_i+\Lambda_i^\theta-P_i)\\
{[\eta']\over\eta}\Bigg|_{z_i} & = & {\kappa^2\over 4}(-3\rho_i-3p_i+
\Lambda_i^\theta-P_i),\nonumber
\eea
where $\Lambda_i$ and $\Lambda_i^\theta$ denote the tensions of usual four dimensional
Minkowski manifold and  $\theta$ direction, correspondingly.
Note how the brane tension, $\Lambda_i$, does not contribute to the
jump of $R'$. The exponential {\it Ansatz} obviously relates the 
energy density on the two branes since $[b']/b|_{z_0}=\pm k=-[b']/b|_{z_1}$, where
the sign in front of $k$ is same as the sign of the difference $z_1-z_0$.

An interesting case may be found if the other brane is assumed to be empty of matter, {\it i.e.}
$\rho_1=p_1=P_1=0$ and even $\Lambda_1^{\theta}=0$. In this case there is only
a four-dimensional cosmological constant at $z_1$ and $\sigma_1=0$. Also we find that
$k=\kappa^2\Lambda_{z=0}$ and , in {\bf  D} -case, the negative cosmological
constant of the bulk reads $\Lambda_B=-\frac 58 \kappa^2\Lambda^2$.

\subsection{Friedmann equations}

The Friedmann equations for the cases {\bf A} and {\bf B} and {\bf  D} can 
be read from the previously obtained solutions (\ref{genefrieds}), again with
$A=k^2$, $B=k$ so, that $\gamma=\sqrt{24\kappa^2\Lambda_B-15k^2}$. 
These special cases of the equations are rather lengthy and not very 
intuitive and hence they are not explicitly presented here.

The Friedmann equations on our brane at $z=z_0$ have interesting
limits when the two branes are well separated from each other. In case {\bf B}
we see that
\bea{6dfrs}
\ Lk\gg 1,\ H^2 & \approx & {\kappa^2\over 4
L}(\rho_0+p_0+\Lambda_0^{\theta}-P_0)-{k\over 4 L}\nonumber\\
 -Lk\gg 1,\ H^2 & \approx & \abs{k}{\kappa^2\over
4}(\rho_0+p_0+\Lambda_0^{\theta}-P_0)
\eea
where we have assumed that $|k|\sim|\sigma_i|$. Hence, in these limit
cases the Friedmann
equation is linear in $\rho_0$ and is not coupled to the matter on the other brane.
Note also that Eqs. (\ref{genefrieds}) are linear in $\sigma_0$ when 
$\sigma_1=0$ recreating the conventional evolution of the scale factor.
In the bowl-model, we again follow the same procedure as before and find
the Friedmann equation on the brane (note, that now we only
have a single brane):
\be{6dEfried}
H^2={\kappa^3\over 2}\sqrt{\Lambda_B}\, \coth(k
z_0)(\rho_0+p_0+\Lambda^{\theta}_0-P_0).
\ee

Finally we wish to give the relations using the observed 4-dimensional quantities.
By using Eq. (\ref{geneenergy}) we can express the pressure $P_{0}$ 
in terms of the $5$-dimensional energy density: By integration we obtain
$\rho_i=3(p_i-P_i)+C$, where $C$ is a constant. 
Moreover, in order to recover the usual Friedmann
equation, we need to express the energy density in terms of the 
four-dimensional quantity. Because the  components of the $5$-dimensional 
energy-momentum tensor are related to the 4-dimensional
counterpart by $(T_A^B)_{4D} = \int dz\,d\theta b(z)\delta(z-z_0)\,\tilde T_A^B$.
We obtain for the exponential models
$(T_A^B)_{4D}=2\pi b(0)e^{kz_0}\tilde T_A^B$ and for the bowl model
$(T_A^B)_{4D}=2\pi b(0)\cosh (kz_0)\tilde T_A^B$. 

Thus the Friedman equations
(\ref{6dfrs}) and (\ref{6dEfried}) may be written in terms of 
4-dimensional quantities.
We find that they are of the form
\be{generic4dFried}
H^2= {8\pi\over 3 M_{eff}^2}(\rho_0 +\Lambda_{eff}),
\ee
where the effective mass scale $M_{eff}$ and effective cosmological constant $\Lambda_{eff}$
depend on the model.
In particular, for the bowl model
\be{Fb2}
{8\pi\over 3 M_{eff}^2 }= {\kappa^3\over 2}{\sqrt{\Lambda_B}\over 2\pi b(0)\sinh(kz_0)}
\ee
and $\Lambda_{eff}=\Lambda^\theta -\frac 13 C$. It is, however, clear, that the effective scale 
$M_{eff}$ is a subject of constraints appearing from standard cosmology,  nucleosynthesis,
recombination {\it etc.} because Hubble rate may change. Therefore 
$M_{eff}$  has to be close enough to $M_{Pl}$.

As the eq. (\ref{generic4dFried}) gives a relation between the parameter of the
model and observed quantities, so does the equation, which determines the strength
of gravitational interaction. Namely, the 4-dimensional gravitational
constant $G_4=M_{Pl}^{-2}$
is now given by
\be{4Dgc}
{1\over G_4} R(z_0,t)^3 = {2\pi \over G_6}\int dz\, {R(z,t)^3 b(z)\over\eta (z,t)},
\ee
where $G_6 = \kappa^2/8\pi =M_*^{-4}$. 
The requirement is thus that the Planck mass $M_{Pl}$ equals now the
usual value observed in the gravitational experiments.

\section{Conclusions}

In this paper we have studied the cosmological evolution of brane
world models with a warped and a number of compact dimensions. By
studying how the bulk evolves with time we find that the induced
evolution on the brane(s) is quite naturally linearly proportional to 
the energy density on the brane. Moreover, in a number cases, the
usual Friedmann equation is reproduced without any dependence 
on the energy density on the so called hidden brane. 

A particularly nice scenario is the bowl model where only one brane is
present and hence there is no hidden brane contribution. The standard
Friedmann equation is recreated with an effective cosmological 
constant on the brane. Interesting models can also be realized by
assuming that the other brane carries only brane tension along the
non-compact dimensions since then the Friedmann equation again has the
usual form. Furthermore, standard cosmological evolution can be
reached in some two-brane models by letting the two branes be well
separated from each other.

An obvious omission in the analysis is the question of radion
stabilization. Here we have to, in addition to the stabilization of
the distance between the brane, be concerned with stabilization of the
extra compact dimensions. Such considerations are clearly needed when
one wishes to fit the models considered here into a larger theoretical
framework. Here our main concern has been to consider these models
from a purely phenomenological point of view and see how cosmological
evolution is affected. In this paper we have also omitted the analysis
of the graviton spectrum in each case which would be needed in order
to study the correction to the Newtonian potential. In the
six-dimensional case this has been done in \cite{ross,me}.

With both warped and, possibly large, compact extra dimensions, one
also needs to keep a number of experimental constraints in mind. The
appearance of the KK-tower of standard model particles sets an upper
limit for the size of the extra compact dimensions on the brane,
$b_i(0)^{-1}>1\TeV$. The corrections to the Newtonian potential must
also be small enough. In order to alleviate the hierarchy problem, the
size of the extra dimensional volume must also be large enough, so
that the effective gravitational constant on our brane is small as
given by Eq. (\ref{4Dgc}). In six dimensions these experimental
constraints were analyzed in detail in \cite{me} and these results are
probably also essentially valid for more than six dimensions. Anyway,
we have found that, at this level, adding compact dimensions surely
increases the number of viable models, despite the new experimental
constraints set by the KK-particle spectrum.

Brane world models with warped and compact dimensions are an
interesting possibility whose implications for e.g. particle 
phenomenology are not yet fully known. Cosmologically they seem 
appealing since the standard
Friedmann evolution with linear dependence on the energy density and
no appearance of hidden brane quantities, is reached in many cases
naturally without any fine tuning.

\section*{Acknowledgements}
This work has been partly supported by the Academy of Finland.

\end{document}